\documentclass[twoside]{dis08}
\usepackage[latin1]{inputenc}
\usepackage[dvips]{graphicx,epsfig,color}
\usepackage{wrapfig,rotating}
\usepackage{amssymb,amsmath,array}
\usepackage{url}

\begin{document}
\title{GPD sum rules: \\ a tool to reveal the quark angular momentum
\footnotetext{
Talk given by D.M.~at DIS 2008, 7-11 April 2008, London  \cite{DM:DIS2008slides}.
}
}

\author{D. M\"uller$^1$, K.~Kumeri{\v c}ki$^2$ and K.~Passek-Kumeri{\v c}ki$^3$
%
%
\vspace{.3cm}\\
%
1- Ruhr-Universit\"at Bochum - Institut f\"ur Theoretische Physik II\\
D-44780 Bochum - Germany
%
\vspace{.1cm}\\
2- University of Zagreb - Department of Physics\\
P.O.B. 331, HR-10002 Zagreb - Croatia\\
\vspace{.1cm}\\
3- Rudjer Bo{\v s}kovi{\'c} Institute - Theoretical Physics Division\\
P.O.Box 180, HR-10002 Zagreb - Croatia\\
}

\maketitle

\begin{abstract}
In deeply virtual exclusive electroproduction  to leading order accuracy one
accesses  generalized parton distributions  on their
cross-over trajectory.  Combining Lorentz covariance and
analyticity leads to a family of GPD sum rules, guiding us to
phenomenological concepts. As an example, we discuss the
constraints from the JLAB/Hall A data on the GPD $E$. Its first
Mellin moment is the anomalous gravitomagnetic moment, which is the
unknown contribution to the quark angular momentum.
\end{abstract}

\section{Introduction}

\phantom{Ref. \cite{DM:DIS2008slides}}
\vspace{-3mm}

Generalized parton distributions (GPDs) might be viewed as {\em
non-diagonal} overlap of  wave functions. This  offers the
opportunity to study the partonic content of the nucleon from a
new perspective. They  are accessible in deeply virtual
leptoproduction of photon and mesons, where the amplitude
factorizes at leading twist-two accuracy in a perturbatively
calculable hard-scattering part and the GPD. Deeply virtual
Compton scattering (DVCS), a subprocess in the leptoproduction of
a photon, is  considered as a theoretically clean process.

One of the main reasons to measure these processes is the quest for
an understanding of the decomposition of the nucleon spin in quark
and gluon angular momenta \cite{Ji96}:
\begin{eqnarray}
\label{SR-AM}
J^Q(\mu^2) + J^G(\mu^2) = \frac{1}{2}\,, \quad
\mbox{with}\quad J^Q(\mu^2)= \sum_{q=u,d,s\cdots}J^q(\mu^2).
\end{eqnarray}
The partonic angular momenta are given by the expectation values
of the corresponding gauge invariant parts of the energy momentum
tensor and might be further decomposed in spin and orbital angular
momenta. They are also expressed by moments of GPDs $H$ and $E$,
\begin{eqnarray}
J^q(\mu^2) = \frac{1}{2} \left[A^q(\mu^2) +B^q(\mu^2)\right], \quad
\left\{ {A^q \atop B^q}\right\}(\mu^2) =
\lim_{\Delta\to0} \int_{-1}^{1}\! dx\, x
\left\{ {H^q \atop E^q}\right\}(x,\eta,t=\Delta^2;\mu^2)\,,
\label{Def-Mom}
\end{eqnarray}
taken in the forward limit, where $\eta \propto \Delta_+$ is the
longitudinal momentum fraction transfer in the $t$-channel. The
quantities $A^q$ are nothing but the averaged momentum fractions
of unpolarized partons They are already phenomenologically
constrained by deeply inelastic scattering measurements. Momentum
conservation guarantees that $A$ is normalized to one and the
angular momentum sum rule (SR) (\ref{SR-AM}) implies then that the
anomalous gravitomagnetic nucleon moment $B$ vanishes:
\begin{eqnarray}
\label{SR-GM}
A\equiv \sum_{q=u,d,s\cdots}A^q(\mu^2) + A^G(\mu^2)=1\,, \quad
B\equiv\sum_{q=u,d,s\cdots}B^q(\mu^2) + B^G(\mu^2) =0\,.
\end{eqnarray}

\section{Setting up the hunting scheme}

Let us first recall the  arguments
that provide an  estimate for the
valence quark%
\footnote{A valence quark is the difference of quark and
anti-quark; the sea is twice the amount of all anti-quarks.}
momentum fractions (\ref{Def-Mom}). The small $x$ behavior of
valence PDFs is governed by the intercept $\alpha \approx 1/2$ of
$\rho$ and $\omega$ Regge trajectories, while  large $x$ counting
rules state a $\sim (1-x)^3$ behavior. Taking also into account
the large $x$ ratio $u/d \sim 5$, the resulting averaged momentum
fractions,
\begin{eqnarray}
\left\{ u_{\rm val} \atop d_{\rm val} \right\}(x,\mu^2)\sim
\frac{35}{32}\frac{(1-x)^3}{\sqrt{x}}
\left\{5 - 3 \frac{99}{80}(1-x)^2 \atop
  1 \right\}
\quad\Rightarrow\quad  A^{u_{\rm val}} \sim 0.32\,,
\quad  A^{d_{\rm val}} \sim 0.11\,,
\end{eqnarray}
are in good agreement with phenomenological findings at a scale
$\mu \sim 2\, {\rm GeV}$.

To estimate the quark angular momenta, we use isospin symmetry to
fix the normalization of $E^{q_{\rm val}}$ in terms of the nucleon
magnetic moments. The relevant Regge intercepts are the same as
before and counting rules state now a $\sim (1-x)^5$ behavior
\cite{Yua03}. The estimates for the anomalous gravitomagnetic
moments and angular momenta follow from Eq.~(\ref{Def-Mom}):
\begin{eqnarray}
B^{u_{\rm val}} \sim 0.13\,, \;\;B^{d_{\rm val}} \sim -0.15
\quad\Rightarrow\quad
J^{u_{\rm val}} \sim 0.2\,,\;\; J^{d_{\rm val}} \sim 0\,.
\label{Est-AGM}
\end{eqnarray}
Hence, the valence part of the anomalous gravitomagnetic  moment
is expected to be small, i.e., $B^{Q}= - B^{G} \sim  B^{\rm sea}$,
and the `unknown' in the spin SR is the sea (or gluon)
contribution.

It was conjectured that $B^{G}$ is zero \cite{Ter99} and so we
would expect that $B^{\rm sea}$  nearly vanishes. In such a
scenario the angular momentum $J^Q \approx A^Q/2 \sim 0.25$ is
essentially expressed by the momentum fraction  and changes only
slightly under evolution. In a covariant two flavor quark model
($B^{G}=0$) one has $B^u=-B^d$, usually within a relatively small
value of $B^{\rm sea}$. In the chiral quark soliton model
($\chi$QSM)  the role of sea quarks is more pronounced and
estimates are compatible with our valence like ones
\cite{Ossetal04}. Lattice measurements are consistent with the
estimates (\ref{Est-AGM}), too. However, because of systematical
errors, including neglecting gluon induced contributions
(disconnected diagrams), we consider $B^{\rm sea}$ as unmeasured.
On the other hand the SR estimate \cite{BalJi97} states that at
least half of the nucleon spin originates from  gluons, i.e.,
$B^{G} > 0$ and $B^{\rm sea}$ is negative. We expect that $|B^{\rm
sea}| \lesssim 1/2$, i.e., $0\lesssim J^Q \lesssim 1/2$.

It is phenomenologically  challenging to reveal the anomalous
gravitomagnetic  moments. So far this has been attempted in a
model dependent way, where the GPD $E^q$ is specifically
parameterized in terms of $J^q$ within a non-Reggeized
`vector-meson exchange' contribution. This flexibility is
incompatible with $\chi$QSM results
\cite{Ossetal04}. Alternatively, one might utilize `common'
GPD models with $B^{\rm sea}$ as a free parameter.

\section{`New' phenomenological tools: GPD sum rules}

The DVCS amplitude is parameterized by Compton form factors
(CFFs). To leading order (LO) accuracy their imaginary parts are
given by GPDs at the {\em cross-over trajectory} $\eta =x$:
\begin{eqnarray}
\label{InvRel}
\Im{\rm m}{\cal F}(\xi,t,{\cal Q}^2) \stackrel{\rm LO}{=}
\pi F^\mp (x=\xi,\eta=\xi,t,{\cal Q}^2)\,, \quad F^-=\{H,E\}\,,\;\;
F^+=\{\widetilde H, \widetilde E\} \,.
\end{eqnarray}
This relation is analogous to the well-known parton interpretation
of DIS structure functions, given as linear combination of PDFs. A
fixed $t$ dispersion relation allows to evaluate the real part of
the CFF from its imaginary part. Hence,  to LO accuracy it is
expressed by the GPD on the cross-over trajectory and a
subtraction constant ($C_{E}=-C_{H}, C_{\widetilde
H}=C_{\widetilde E}=0$):
\begin{eqnarray}
\Re{\rm e}{\cal F}(\xi,t,{\cal Q}^2) \stackrel{\rm LO}{=}
 {\rm PV}\int_{0}^{1}\!
dx\left(\frac{1}{\xi-x} \mp
\frac{1}{\xi+x} \right) F^\mp (x,x,t,{\cal Q}^2) +C_{F}(t,{\cal Q}^2)\,.
\label{Def-DisRel1}
\end{eqnarray}
The scale dependence of a CFF is governed by the GPD in the outer
region ($\eta \le x$) and that radiative corrections extend
Eq.~(\ref{InvRel}) to a convolution integral over the outer
region.

The CFFs can be evaluated without knowing the GPDs in the central
region. Combining operator product expansion and dispersion
relation shows that the GPD in this  region arises from the
Lorentz covariant decomposition of the CFFs in terms of operator
matrix elements \cite{KumMuePas07}. This can be also derived from
a integral equation \cite{Ter05,DieIva07}, denoted as {\em GPD SR
family}:
\begin{eqnarray}
\label{Def-GDPSR}
\int_0^{1}\! dx\;
\left(\frac{1}{\xi-x}\mp\frac{1}{\xi+x}\right)
\left[F^{\mp}(x,\eta=\vartheta \xi,t,Q^2)-F^{\mp}(x,\eta =
\vartheta x,t,Q^2)\right]=C_F(\vartheta,t,Q^2)\,.
\end{eqnarray}
The application of the GPD  SR  family  (\ref{Def-GDPSR}) is
manifold \cite{KumMuePas08}. They allow us to construct the GPD in
the central region from its knowledge in the outer region and the
subtraction constant. We note that the subtraction constant is
entirely related to the so-called $D$-term and only contributes to
the $t-$channel  $J=0$ angular momentum contribution. In analogy
to finite energy SRs,  one might write down GPD ones that connect
the `low' and `high' energy content of the GPD.

Additionally, one might set up a  GPD model on its cross-over
trajectory by factorizing it into a GPD for $\eta=0$ and a
skewness function $S(x,t,{\cal Q}^2|F^{\mp})$:
\begin{equation}
\label{Def-skePar}
 F^{\mp}(x,x,t,{\cal Q}^2) = \Big[1+ S(x,t,{\cal
Q}^2|F^{\mp})\Big] F^{\mp}(x,\eta=0,t,{\cal Q}^2)\,.
\end{equation}
Taking the limit $\xi\to 0$ in Eq.~(\ref{Def-GDPSR}),
one finds a constraint for the skewness function
\begin{eqnarray}
\label{Sum-Rul-DVCS-0} \int_{(0)}^{1}\! dx\; \frac{1}{x}
S(x,t,{\cal Q}^2|F^-) F^{-}(x,\eta=0,t,{\cal Q}^2) =\frac{1}{2}{
C}_F(t,{\cal Q}^2) \,, \qquad
\end{eqnarray}
where $(0)$ indicates analytic regularization.

The skewness function can be simply evaluated within a given GPD
model. To make contact with both phenomenology and lattice
measurements, it is more appropriate to consider Mellin moments.
The skewness effects might be quantified by deviation factors:
\begin{eqnarray}
\label{Cal-del} \delta_j(t,\mu^2 | F^\mp)   =
 \frac{\int_{0}^1\! dx\, x^j
S(x,t,\mu^2|F^\mp)F^\mp(x,\eta=0,t,\mu^2)}{\int_{0}^1\! dx\, x^j
F^\mp(x,\eta=0, t,\mu^2)} = \sum_{\substack{n=2 \\ {\rm
even}}}^\infty \frac{f_{j+n}^{(n)}(t,\mu^2)}{ {f}_j(t,\mu^2)}\,.
\end{eqnarray}
They are given by a series of local operator matrix elements
$f_{j+n}^{(n)}(t,\mu^2)$ with spin $j+n+1$, containing $n$ total
derivatives. The state of the art in lattice measurements is the
evaluation of spin-three operator matrix elements, allowing for a
first guess of $\delta_0(\cdots|H) \sim 0.2$.

\section{Accessing GPD $E$ from experimental data}

To hunt for the anomalous gravitomagnetic
moment one should read formula (\ref{Cal-del}) as \cite{KumMuePas08}
\begin{eqnarray}
B({\cal Q}^2) \equiv e_1(t=0,{\cal Q}^2)\stackrel{\rm LO}{=} \frac{1}{\pi} \frac{
\int_{0}^1\! d\xi\, \xi\, \lim_{t\to 0}\Im{\rm m} {\cal E}(\xi, t,{\cal
Q}^2)}{1+\delta_1(t=0,{\cal Q}^2 |E^-)}\,.
\label{Acc-AM}
\end{eqnarray}
Certainly, it will be challenging to measure $\Im{\rm m} {\cal
E}$. We emphasize that a measurement of the real and imaginary
part allows to utilize  the `dispersion' integral
(\ref{Def-DisRel1}) as a SR.  Assuming a Regge-like extrapolation
in the small $\xi$ region,   one might then even  extract $\Im{\rm
m} {\cal E}$ in the large $\xi$ region. The deviation factor
$\delta_1$ in Eq.~(\ref{Acc-AM}) reminds us that a `measurement'
of the anomalous gravitomagnetic moment requires an understanding
of the skewness effect.

The CFF $\cal E$ might be measured in DVCS on neutron, from the
transverse proton spin asymmetry or  the beam charge asymmetry at
small $\xi$. Presently, one can only obtain a `local' constraint
on the GPD $E$. We have modelled valence GPDs at $\eta=0$,
adjusted them to the nucleon form factors and PDFs. We have varied
the skewness function, constrained by Eq.~(\ref{Def-skePar}) and
JLAB/Hall A DVCS off proton data \cite{Cametal06}, and  estimated
the  sea quark contribution to $H$ by a GPD model dependent
extrapolation of DVCS measurements in collider kinematics. We
have  found that  $H$ and $\widetilde H$ contribute only little to the
interference term in DVCS off neutron  \cite{Mazetal07} and that
the $E$-constraint is mainly given by the experimental error
($\Delta^{\rm exp} \approx \pm 0.5$):
\begin{eqnarray}
 \left| E^{u_{\rm val}} +  4 E^{d_{\rm val}} +  2 E^{\rm sea}
\right|(\xi,\xi, t )\Big|_{ {\xi= 0.22 \hfill\atop t \approx -
0.4\, {\rm GeV}^2 } }  \lesssim \frac{9}{\pi} \frac{4 M^2}{-t}
\frac{ \left|\Delta^{\rm exp} \right|}{F_2(t)} \Big|_{ t \approx -
0.4\, {\rm GeV}^2 } \approx 20\,.
\label{Con-E}
\end{eqnarray}
With our  ansatz for $E$ and supposing $\delta_0 =0.2$ we found it
likely that the valence contribution in (\ref{Con-E})
is about $-5 \pm 2$.  Being optimistic, we might state
that $E^{\rm sea}$ is constrained by
$$-7  \lesssim E^{\rm
sea}(\xi= 0.22, t \approx - 0.4\, {\rm GeV}^2 ) \lesssim 13. $$
This is a rather weak condition, since for this kinematical point
the modulus $|E^{\rm sea}|$ should be in a pessimistic case of a
large skewness effect restricted to be $\lesssim 10 |B^{\rm
sea}|$. This  exceeds the interval of $|B^{\rm sea}| \lesssim
0.5$, which covers $0 \lesssim J^Q \lesssim 1/2$.\\

\noindent We conclude  that dispersion techniques should be
employed to reveal GPDs on its cross-over trajectory from present
DVCS measurements in fixed target kinematics. This might lead to a
better GPD understanding,  needed to access the quark angular
momentum from dedicated experiments.

\begin{footnotesize}

\end{footnotesize}

\end{document}